\documentclass[journal,twocolumn]{IEEEtran}

\usepackage{array}
\usepackage[caption=false,font=normalsize,labelfont=sf,textfont=sf]{subfig}
\usepackage{stfloats}
\usepackage{url}
\usepackage{verbatim}
\usepackage{graphicx}
\def\BibTeX{{\rm B\kern-.05em{\sc i\kern-.025em b}\kern-.08em
    T\kern-.1667em\lower.7ex\hbox{E}\kern-.125emX}}

\usepackage{cite}
\usepackage{amsmath,amssymb,amsfonts}
\usepackage{algorithmic}
\usepackage{graphicx,color}
\usepackage{booktabs}
\usepackage{textcomp}
\usepackage{hyperref}
\usepackage{siunitx}
\DeclareSIUnit{\MVA}{MVA}
\usepackage{capt-of}
\usepackage{microtype}

\newlength{\colw}
\newlength{\colsep}
\newcommand{\field}[2]{%
\par\addvspace{2pt}\noindent \hangindent=\dimexpr\colw+\colsep\relax\hangafter=1 \makebox[\dimexpr\colw+\colsep\relax][l]{\ttfamily\bfseries #1}#2\par}

\begin{document}
\title{A Benchmark Graph Dataset for Transient Stability Assessment of the IEEE 9-Bus System: 20,000 Scenarios with Full Generator Trajectories}
\author{Hussein Suprême,~\IEEEmembership{Senior Member,~IEEE,} Martin de Montigny and Arnaud Zinflou,~\IEEEmembership{Senior Member,~IEEE}
\thanks{Manuscript created August, 2026. Dr. Hussein Suprême, Dr. Martin de Montigny and Dr. Arnaud Zinflou are with Institut de recherche d’Hydro-Québec, Varennes, QC, Canada, J3X 1S1. Corresponding author e-mail: supreme.hussein2@hydroquebec.com).}}

\markboth{Institut de recherche d'Hydro-Québec (IREQ), August~2026}%
{How to Use the IEEEtran \LaTeX \ Templates}

\maketitle

\begin{abstract}
Transient stability assessment determines whether a power system retains synchronism after a large disturbance. Machine-learning surrogates can accelerate it, but progress is limited by the lack of open datasets that combine dynamic ground truth, network graph structure, and machine parameters. We release a benchmark of 20,000 three-phase-to-ground fault scenarios on the IEEE 9-bus system. Each scenario couples an AC power-flow operating point with a detailed electromagnetic-transient simulation of the post-fault response. Every record provides the network as an attributed graph (nine buses, eighteen directed branches, ten node and twelve edge features), the full rotor-angle and speed trajectories of the three generators, the static machine constants, the fault description, and a center-of-inertia binary stability label. Wide load and generation scalings across eighteen fault locations yield a near-balanced distribution (48.96\% stable, 51.04\% unstable). Generation is deterministic and fully reproducible through fixed seeds and public code. The dataset is distributed on IEEE DataPort under a persistent \href{https://ieee-dataport.org/documents/10000-scenario-graph-dataset-full-generator-trajectories-transient-stability-assessment}{DOI} and supports stability classification, trajectory prediction, margin and critical-clearing-time estimation, and the comparison of topology-aware, physics-based, and hybrid learning methods.
\end{abstract}


\section{Background}
\IEEEPARstart{T}{ransient} stability assessment (TSA) evaluates whether the synchronous machines of a power system remain in synchronism following a large disturbance such as a short circuit~\cite{9286772}. The reference method is time-domain simulation of the electromechanical dynamics, which is accurate but computationally expensive, motivating machine-learning surrogates that infer stability directly from the operating point and disturbance~\cite{7726772}. 

Two modeling families dominate recent work. Graph neural networks (GNNs) exploit the fact that a power network is naturally a graph, propagating information along the electrical topology and generalizing across operating points that share the same structure. Physics-informed neural networks (PINNs) instead embed the governing swing-equation dynamics into the training objective, improving data efficiency and physical consistency. Under the classical machine model, the motion of generator $i$ obeys:

\begin{equation} 
\frac{2H_i}{\omega_s}\,\frac{d\omega_i}{dt} + D_i\,\omega_i = P_{m_i} - P_{e_i}(\boldsymbol{\delta}) \label{eq:swing} 
\end{equation}
where $\delta_i$ is the rotor angle, $\omega_i = d\delta_i/dt$ the rotor speed deviation, $H_i$ the inertia constant, $D_i$ the damping coefficient, $\omega_s$ the synchronous speed, $P_{m_i}$ the mechanical power, and $P_{e_i}$ the electrical power output. A PINN penalizes the residual of \eqref{eq:swing}, so evaluating it requires the machine constants ($H_i$, $D_i$) together with the per-generator trajectories ($\delta_i$, $\omega_i$), while the mechanical power is set to its pre-fault value $P_{m_i} = P_{g_i}$ in first-swing analysis. A hybrid physics-informed graph model (PIGNN) that combines topological message passing with this swing-equation residual is a natural next step, but a controlled comparison of the three approaches requires a dataset that simultaneously exposes (i) the network graph, (ii) the physical machine constants needed to evaluate~\eqref{eq:swing}, and (iii) the full post-fault trajectories that serve as dynamic ground truth. 

Existing open TSA resources typically provide only scalar stability labels or omit the machine parameters and per-generator trajectories required by physics-informed methods. The primary contribution of this dataset is therefore not network scale, but annotation richness: the simultaneous availability of graph topology, physical machine parameters, complete generator trajectories, deterministic reproducibility, and public accessibility within a single benchmark.

\section{Selection Methods and Design}
\subsection{Base system}
The base network is the IEEE 9-bus, three-machine test system operating at \SI{60}{\hertz} with a \SI{100}{\MVA} base. The system is originally defined as a Simscape Electrical model~\cite{mathworks_ieee9bus_loadflow}, from which the network topology, branch parameters, and synchronous-machine data were converted into an equivalent MATPOWER case~\cite{simscape2matpower}. This dual representation underpins the two-stage generation pipeline: the MATPOWER case serves as the steady-state engine that produces the pre-fault operating point, which is then mapped back to the Simscape Electrical model to run the dynamic simulation.The three synchronous machines are represented using the classical model, and their static constants are held fixed across all scenarios (Table~\ref{tab:machines}). To avoid redundancy, these constants are stored once with the dataset rather than duplicated in each record, and are re-attached to every sample at load time. The damping coefficient is set to zero, since first-swing transient stability is governed primarily by the accelerating energy imparted during the fault, for which the damping contribution is typically negligible.

\begin{table}[ht] 
\caption{Static Parameters of the Synchronous Machine Models.} \label{tab:machines} 
\begin{tabular*}{\linewidth}{@{\extracolsep{\fill}}lccc} 
\toprule Quantity & Gen~1 & Gen~2 & Gen~3 \\ 
\midrule Inertia $H$ (s) & 23.64 & 6.40 & 3.01 \\ 
Damping $D$ & 0 & 0 & 0 \\
$X_d$ (pu) & 0.146 & 0.8958 & 1.3125 \\
$X_d'$ (pu) & 0.0608 & 0.1198 & 0.1813 \\ 
Rating (MVA) & 247.5 & 192 & 128 \\
\bottomrule 
\end{tabular*} 
\end{table}

\subsection{Graph-based representation of operating conditions}
Each operating condition is represented as a directed attributed graph: 
\begin{equation} 
\label{eq:directed_graph}
\mathcal{G} = (\mathcal{V}, \mathcal{E}, \mathbf{X}, \mathbf{E}), \end{equation} 
where $\mathcal{V}=\{1,\ldots,N\}$ denotes the set of buses and $\mathcal{E}\subseteq \mathcal{V}\times\mathcal{V}$ denotes the set of directed branches. 
Each physical transmission element connecting buses $i$ and $j$ is represented by two directed edges, $(i,j)$ and $(j,i)$, allowing branch quantities to be consistently associated with sending and receiving ends. The network topology is described by the adjacency matrix $\mathbf{A}\in\{0,1\}^{N\times N}$, where: 
\begin{equation} 
\label{eq:adj_mat}
A_{ij}= \begin{cases} 1, & (i,j)\in\mathcal{E},\\ 0, & \text{otherwise}. \end{cases} 
\end{equation} 
or equivalently by the edge-index matrix $\mathbf{I}\in\mathcal{V}^{2\times M}$, whose $k$-th column $(s_k,r_k)$ identifies the sending and receiving buses of edge $k$. 
Node features are collected in~\eqref{eq:node_feat}, where $F_v$ denotes the dimension of the node feature vector in~\eqref{eq:node_feat_vect}. Each vector contains the bus type~$\tau_i$, nominal voltage level~$\kappa_i$, generator identifier~$g_i$, load identifier~$l_i$, active~$P^d_{i}$ and reactive~$Q^d_{i}$ demands, voltage magnitude~$V_i$ and angle~$\theta_i$, and active~$P^g_{i}$ and reactive power~$Q^g_{i}$ generation obtained from the pre-fault power-flow solution.
\begin{equation} 
\label{eq:node_feat}
\mathbf{X} = \left[\mathbf{x}_1,\ldots,\mathbf{x}_N\right]^\top \in \mathbb{R}^{N\times F_v}
\end{equation}
\begin{equation} 
\label{eq:node_feat_vect}
\mathbf{x}_i= \left( \tau_i,\, \kappa_i,\, g_i,\, \ell_i,\, P_i^d,\, Q_i^d,\, V_i,\, \theta_i,\, P_i^g,\, Q_i^g \right) \end{equation} 

Similarly, edge features are gathered in~\eqref{eq:edge_feat}, where the feature vector associated with edge $k=(i,j)$ is given by~\eqref{eq:edge_features}.
\begin{equation} 
\label{eq:edge_feat}
\mathbf{E} = \left[\mathbf{e}_1,\ldots,\mathbf{e}_M\right]^\top \in \mathbb{R}^{M\times F_e}
\end{equation}

\begin{equation}\label{eq:edge_features}
\begin{aligned} \mathbf{e}_k = (&R_{ij}, X_{ij}, B_{ij}, t_{ij}, \phi_{ij}, S_{ij}^{\max}, \\ &u_{ij}, P_{ij}^{f}, Q_{ij}^{f}, P_{ij}^{t}, Q_{ij}^{t}, \lambda_{ij}) 
\end{aligned}
\end{equation}

The edge attributes include the network parameters, namely the series resistance $R_{ij}$, reactance $X_{ij}$, shunt susceptance $B_{ij}$, off-nominal tap ratio $t_{ij}$, phase-shifting angle $\phi_{ij}$, thermal limit $S_{ij}^{\max}$, and line status $u_{ij}$, together with the operating-point-dependent quantities obtained from the power-flow solution: sending-end power flow $(P_{ij}^{f},Q_{ij}^{f})$, receiving-end power flow $(P_{ij}^{t},Q_{ij}^{t})$, and the normalized line loading~\eqref{eq:norm_line_loading}.
\begin{equation} 
\label{eq:norm_line_loading}
\lambda_{ij} = \frac{|S_{ij}|}{S_{ij}^{\max}} 
\end{equation} 

The network topology, branch parameters, and machine data remain identical across all scenarios. Consequently, the variability between scenarios is entirely captured through the operating-point-dependent node and edge features, namely $(V_i,\theta_i,P_i^g,Q_i^g,P_i^d,Q_i^d)$ and $(P_{ij}^{f},Q_{ij}^{f},P_{ij}^{t},Q_{ij}^{t},\lambda_{ij})$, which characterize the pre-disturbance system state.

\subsection{Two-stage generation pipeline}
\textbf{Steady state.} A randomized operating point is created by scaling bus loads by a factor drawn uniformly from $[0.60,\,1.25]$ and generator dispatch by a factor drawn uniformly from $[0.90,\,1.10]$. An AC power-flow (PF) solution (MATPOWER~8.1)~\cite{matpower81} establishes the pre-fault voltages, angles, injections, and branch flows. 
To ensure a physically realistic active-power output at the slack bus, a dispatch resulting from the PF solution is verified against its capability limits $[P_{\min},\,P_{\max}]$. If the slack output falls below $P_{\min}$ (including negative values) or exceeds $P_{\max}$, the corresponding active-power surplus or deficit is redistributed among the non-slack generators in proportion to their available downward or upward margins, i.e., $P_{g}-P_{\min}$ or $P_{\max}-P_{g}$, respectively. The slack generator is then reset within its admissible range and the PF is recomputed to restore network-wide power balance. This margin-based correction guarantees that every retained operating point respects generator capability limits and avoids unrealistic reference-bus injections.
Each steady-state solution is screened on three physical criteria and assigned one of three quality tags. Bus voltages are evaluated on load 
buses only, since generator buses hold their scheduled setpoint. The tags are defined in Table~\ref{tab:opf_classification}.


\begin{table}[ht] 
\caption{Operating-Point Acceptance Criteria.} \label{tab:opf_classification} 
\begin{tabular*}{\linewidth}{@{\extracolsep{\fill}}lp{0.72\linewidth}} 
\toprule Class & Description \\
\midrule \textsc{Accepted} & PF converged; $V_{\mathrm{load}}\in[0.94,1.06]$ pu; no generator Q-limit violation; maximum line loading $\leq 90\%$. \\
\textsc{Borderline} & PF converged but at least one margin exceeded: Q-limit violation, line loading $>90\%$, or $V_{\mathrm{load}}\notin[0.94,1.06]$ pu and $V_{\mathrm{load}}\in[0.90,1.10]$ pu. \\ 
\textsc{Rejected} & PF non-convergent or $V_{\mathrm{load}}\notin[0.80,1.20]$ pu. Excluded from dynamic simulations. \\ 
\bottomrule 
\end{tabular*} 
\end{table}

\noindent \textbf{Transient response.} The accepted operating point initializes a detailed electromagnetic-transient (EMT) model of the machines and network. A three-phase-to-ground (3PHG) fault is applied at \SI{1.0}{\second} at one of fifteen locations (each of the nine buses, or the mid-point of each of the nine lines) and cleared after a duration drawn uniformly from $[0.080,\,0.450]$~\si{\second}. The simulation runs to \SI{3.0}{\second} with an \texttt{ode23tb} solver at a maximum step of
\SI{1e-3}{\second}, recording the rotor angle and speed of every generator.

\begin{figure}[ht]
\centering \includegraphics[width=\columnwidth]{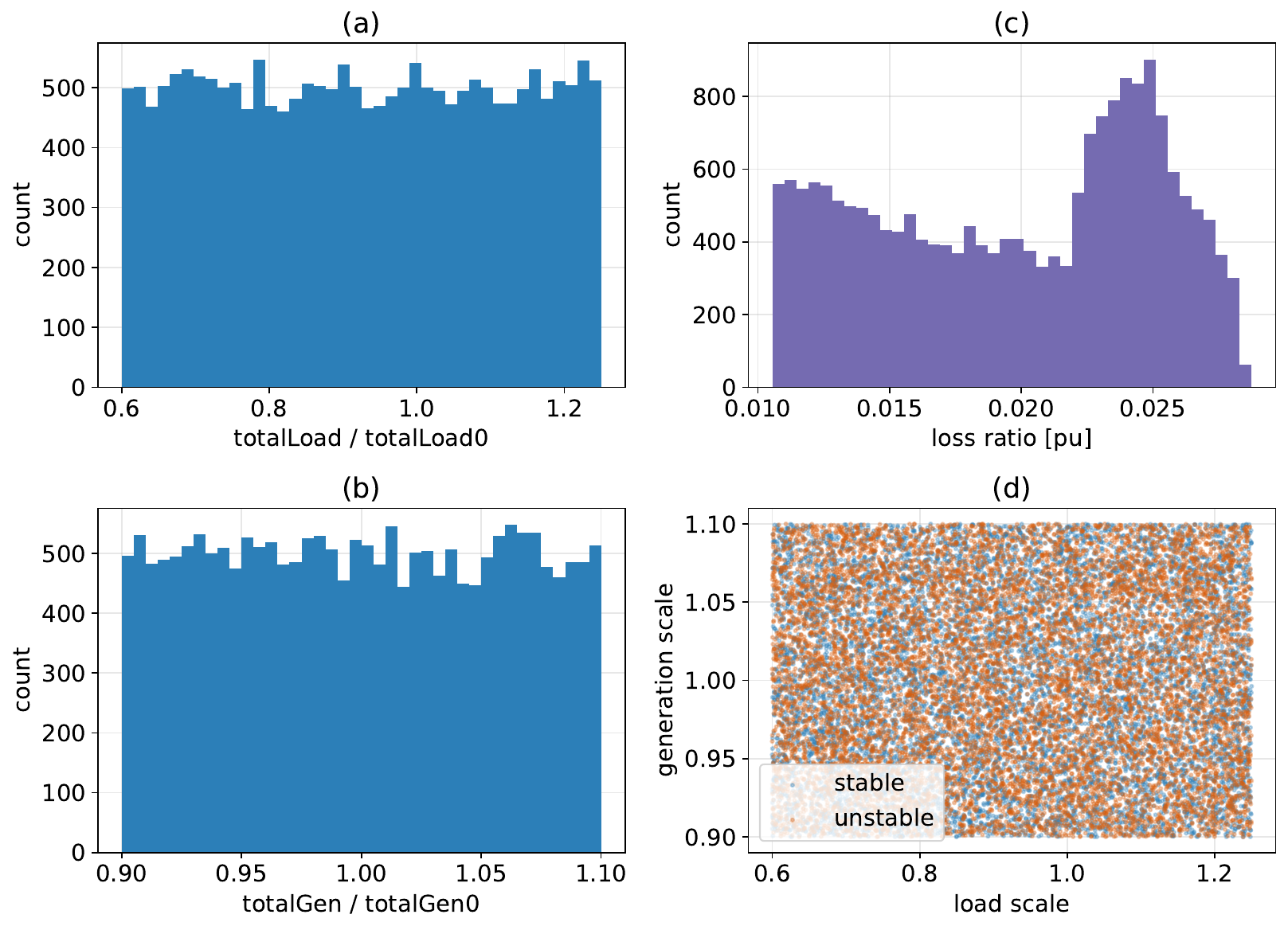} \caption{Quantitative characterization of the sampled operating conditions. No capability-limit violations occur in this dataset. (a) Distribution of the load-scaling factor; (b) distribution of the generation-scaling factor; (c) distribution of the resulting network loss ratio (pu), increasing with system loading; and (d) joint distribution of scenarios in the load--generation scaling plane, colored by stability outcome.} 
\label{fig:scaling_loss} 
\end{figure}

Figure~\ref{fig:scaling_loss} quantifies the operating-point diversity of the dataset. Panels (a) and (b) confirm that both scaling factors are sampled with approximately uniform coverage across their ranges, so the dataset exercises light- and heavy-loading conditions rather than clustering near the nominal point. The resulting loss ratio in panel (c) increases monotonically with system loading, consistent with the quadratic dependence of transmission losses on current. In the load--generation plane of panel (d), stable scenarios concentrate at lower scalings while unstable ones dominate the high-stress region, separated by a broad overlap band. This overlap confirms that stability is not trivially separable from the scaling factors alone, so the dataset provides genuinely informative, non-degenerate operating conditions for learning.
\subsection{Stability label}
For each transient run, the center-of-inertia (COI) reference is computed and the maximum pairwise rotor-angle separation over the simulation horizon, $\Delta_{\max}$ (in degrees), is extracted~\cite{7776913}. A scenario is labeled \textsc{unstable} if $\Delta_{\max} > 180^{\circ}$ and \textsc{stable} otherwise. This threshold is a modeling choice; the separation is stored per record so that alternative thresholds can be applied without re-simulation.

\section{Validation and Quality}
\begin{figure}[ht]
\centering
\includegraphics[width=\columnwidth]{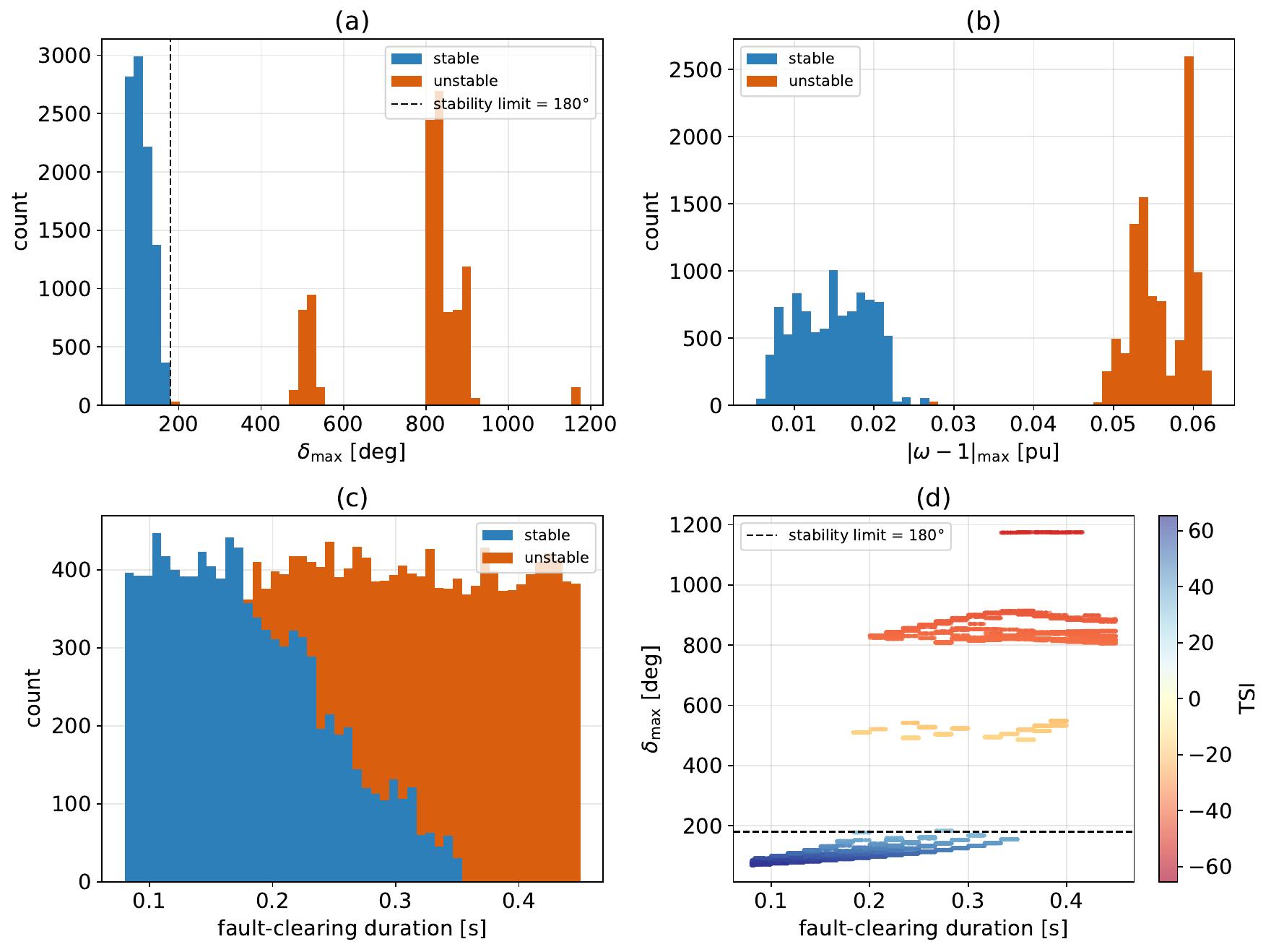}
\caption{Physics-based validation of the dynamic layer.
(a)~Maximum COI rotor-angle separation with the $180^{\circ}$ first-swing
stability limit (the dataset's decision boundary); (b)~maximum rotor-speed
deviation, both coloured by class; (c)~fault-clearing duration by class; and
(d)~clearing duration versus $\Delta_{\max}$ (coloured by TSI): the stable
branch grows smoothly below the $180^{\circ}$ limit, and the transition to
instability occurs in the $0.18$--$0.35$~s range.}
\label{fig:physics_panels}
\end{figure}

Figure~\ref{fig:physics_panels} presents four physics-based consistency
checks that validate the dynamic layer of the dataset and confirm that the
stability labels are physically meaningful rather than artefacts of the
labelling procedure. Panels~(a) and~(b) report the two state variables of
the classical machine model, the maximum COI rotor-angle separation
$\Delta_{\max}$ and the maximum rotor-speed deviation $\max_t|\omega-1|$.
Both quantities are strongly bimodal and cleanly separated by class: stable
scenarios remain below the $180^{\circ}$ first-swing stability limit
(reaching at most $177^{\circ}$, with speed deviations in the range
$0.005$--$0.027$~pu), whereas unstable scenarios lie above it, starting
right at the limit ($183^{\circ}$ for a small marginal group) and forming a
dense cluster from $\approx\!486^{\circ}$ upward (speed deviations of
$0.028$--$0.062$~pu). The two modes are separated by a wide empty band, with
no scenario in the interval $[178^{\circ},\,183^{\circ}]$ and none between
$184^{\circ}$ and $486^{\circ}$, and the angle- and speed-based descriptors
are mutually consistent, providing an independent cross-validation of the
binary label.

The large separation values observed for unstable cases (up to
$\sim\!1175^{\circ}$, i.e.\ more than three revolutions relative to the
center of inertia) are the expected consequence of the undamped classical
model: once synchronism is lost, the rotor angle grows without bound over the
$3.0$~s integration horizon. Accordingly, $\Delta_{\max}$ is used only to
derive the binary first-swing label and is not interpreted as a continuous
physical magnitude; the label is assigned as soon as the separation first
exceeds the limit.

Panels~(c) and~(d) examine the causal relationship between the fault-clearing
duration and the stability outcome. In panel~(c), unstable scenarios exhibit
systematically longer clearing durations (median $\approx 0.35$~s) than stable
ones (median $\approx 0.17$~s), consistent with the accumulation of
accelerating energy during the fault. The partial overlap of the two
distributions (stable cases up to $\approx 0.35$~s, unstable cases down to
$\approx 0.18$~s) is expected and desirable: because the critical clearing
time depends on the fault location, a given clearing duration can be stable at
one site and unstable at another. Panel~(d) makes this mechanism explicit.
Along the stable branch, $\Delta_{\max}$ increases smoothly and almost
monotonically with the clearing duration while remaining below the
$180^{\circ}$ limit, and the transition to instability occurs in a narrow band
around $0.18$--$0.35$~s, which coincides with the critical clearing time of the
IEEE 9-bus system reported in the literature (typically $0.2$--$0.35$~s for
three-phase faults)~\cite{SULISTIAWATI2016345}. The discrete horizontal bands
of unstable points (near $520^{\circ}$, $830^{\circ}$, and $1170^{\circ}$)
correspond to distinct loss-of-synchronism modes associated with different
machine groups and fault locations. Together, these four panels confirm that
the dataset spans a physically realistic operating range, that the two
stability criteria agree, and that the clearing duration and fault location
carry genuine discriminative information, making them meaningful stratification
variables for a train/test split.

The binary label is the standard first-swing criterion: a 
scenario is declared
unstable when the peak inter-machine separation exceeds the $180^{\circ}$ limit
shown in panel~(a) (dataset field \texttt{transient.label.value}, with
\texttt{threshold}~$=180^{\circ}$), which reproduces the reference stable count
of $9{,}762$ scenarios ($48.81\%$) exactly. The transient stability index (TSI),
used in panel~(d) as a continuous colour scale,

\begin{equation}
\label{eq:tsi}
\mathrm{TSI} = \frac{360^{\circ}-\Delta_{\max}}{360^{\circ}+\Delta_{\max}}
\times 100,
\end{equation}
provides a compatible margin. Because no scenario falls in the interval
$[184^{\circ},\,486^{\circ}]$, the labelling is essentially insensitive to the
exact threshold: adopting the $360^{\circ}$ ($\mathrm{TSI}<0$) convention
instead of the $180^{\circ}$ first-swing limit would reclassify only $30$ of
$20{,}000$ scenarios ($0.15\%$), all marginal cases with~$180^{\circ}<\Delta_{\max}\le 183^{\circ}$.

\begin{figure}[ht]
\centering
\includegraphics[width=\columnwidth]{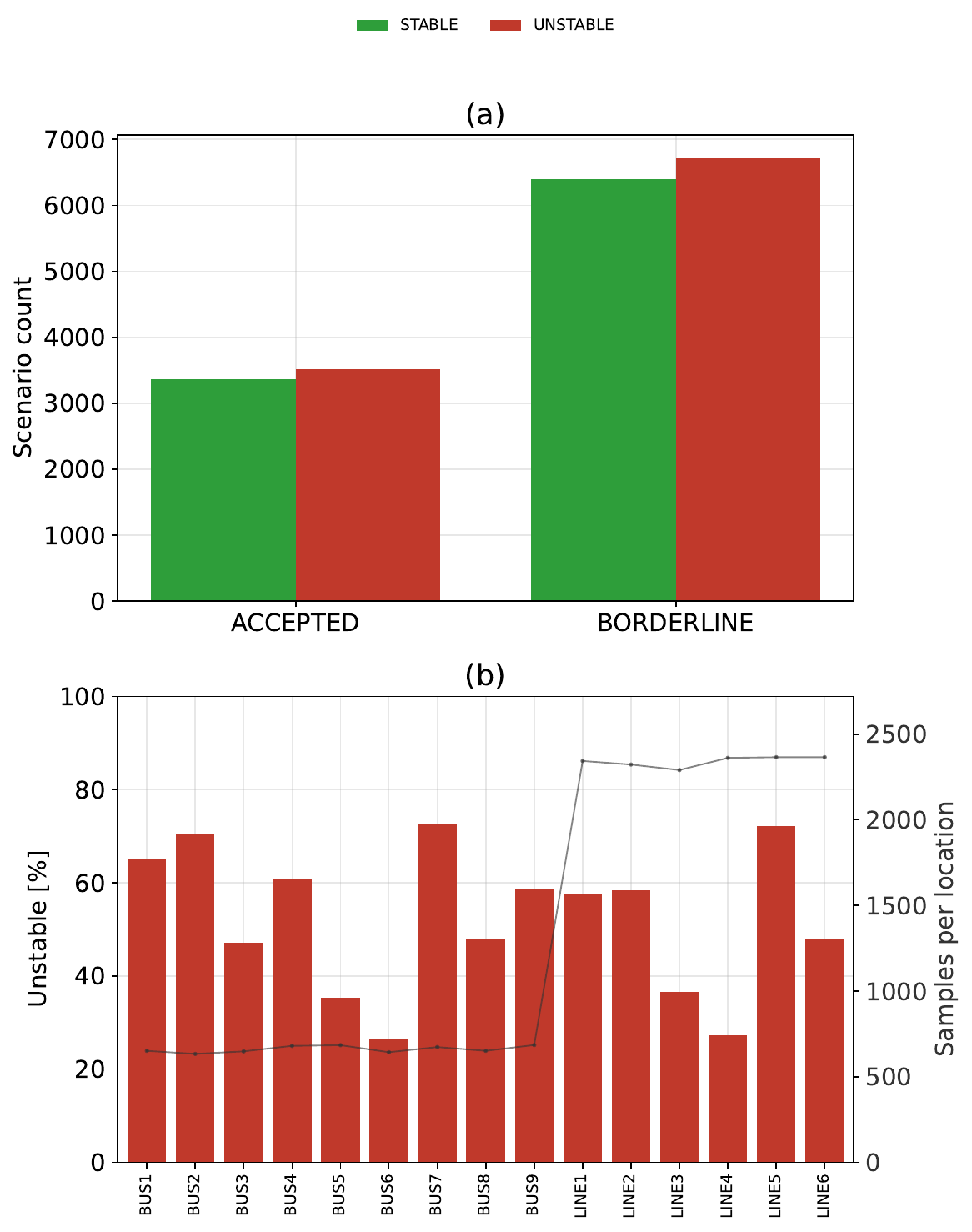}
\caption{Composition cross-checks: (a) Steady-state screening status against the transient stability label; (b) Per-fault-location unstable rate.}
\label{fig:status_location}
\end{figure}
Figure~\ref{fig:status_location} reports two composition cross-checks that validate the design. Panel~(a) contrasts the steady-state screening status with the transient stability label. Crucially, both statuses contain both classes: the \textsc{accepted} and \textsc{borderline} operating points each split into stable \emph{and} unstable outcomes ($3,372/3,501$ and $6,420/6,707$, respectively), so a feasible pre-fault point is in no way a guarantee of post-fault synchronism. This confirms that the pre-fault feasibility tag is not a proxy for the stability outcome: the two carry distinct information and must be kept as separate fields, the tag being a data-quality descriptor rather than a shortcut to the target. Based on Table~\ref{tab:opf_classification}, any operating-point was rejected.
Approximately 35\% of the unstable cases lost synchronism before fault clearing. Since TSA is primarily concerned with post-fault system behavior, these cases are not considered physically meaningful. They can be identified through the variable \texttt{time2LossSync}, defined as the duration between fault clearing and loss of synchronism, with negative values indicating that synchronism was lost during the fault.

Panel~(b) reports the per-fault-location unstable rate for each of the fifteen fault sites that carry faults, excluding the transformers branches. For a location $\ell$, it is defined as the percentage of that location's labelled scenarios whose label is \textsc{unstable}, 
\begin{equation} \rho_{\ell} \;=\; 100\times \frac{\#\{\, k : \mathrm{loc}(k)=\ell \;\wedge\; y_k = 1 \,\}} {\#\{\, k : \mathrm{loc}(k)=\ell \,\}} \quad [\%] 
\label{eq:unstable_rate} 
\end{equation} 
where $y_k\in\{0,1\}$ is the stability label ($1=$ unstable), and only scenarios with a valid transient label and a defined location contribute to both numerator and denominator. The rate varies markedly across sites, from $27.3\%$ (\textsc{line4}) to $72.7\%$ (\textsc{line5}), so the fault location carries genuine discriminative signal rather than trivially producing all-stable or all-unstable cases, making it a meaningful stratification variable for the train/test split.

\section{Records and Storage}
The dataset~\cite{d10j-5b27-26} is archived under a persistent DOI on IEEE DataPort and distributed as a single MATLAB \texttt{.mat} file (v7.3/HDF5, $\approx 2.85$~GB) containing a cell array \texttt{dataset.samples\{k\}} of per-scenario records together with \texttt{dataset.statistics} and metadata. Static machine constants and the reference partition are provided as small companion files (\texttt{machine\_parameters.mat}, \texttt{dataset\_split.mat}). Because the container is HDF5-based, records are readable from MATLAB, Python (\texttt{h5py}/\texttt{scipy}), and other HDF5 clients, and a fixed-grid export to per-scenario HDF5 graphs is supported for machine-learning pipelines. Field names, units, and feature groups are embedded in each record.

\subsection{Record structure}
Each scenario is stored as a single MATLAB structure (\texttt{dataset.samples\{k\}}) comprising nine top-level fields. The design deliberately separates the \emph{steady-state feasibility status} (\texttt{validation}) from the \emph{transient-stability label} (\texttt{transient.label}): the two answer different questions and are never conflated. Table~\ref{tab:fields} summarizes the fields. Static synchronous-machine constants are shared by all scenarios; rather than being duplicated per sample, they are stored once in a companion file and re-attached to each record at load time, indexed by \texttt{genNames}.
\\

\captionof{table}{Top-level fields of each scenario record \texttt{dataset.samples\{k\}}.} \label{tab:fields}
\noindent\rule{\linewidth}{0.8pt}\par\addvspace{2pt} \noindent\makebox[\dimexpr\colw+\colsep\relax][l]{\textbf{Field}}\textbf{Description}\par \addvspace{2pt}\noindent\rule{\linewidth}{0.4pt}
\field{metadata}{Provenance of the sample: base case name (\texttt{caseName}), scenario identifier (\texttt{scenarioID}), and creation timestamp.} 
\field{fault}{Disturbance definition: fault type, location (\texttt{locationType} \textsc{bus}/\textsc{line} with the corresponding \texttt{bus} or \texttt{line} index), fault start time, clearing time, and duration.} 
\field{powerFlow}{Pre-fault operating point from the converged AC PF solution: bus voltage magnitude $V$ and angle $\theta$, bus active and reactive demand $P_d$/$Q_d$, and generator active and reactive output $P_g$/$Q_g$.} 
\field{scenario}{Sampled perturbations applied to the base case to generate this operating point, namely the load- and generation-scaling draws and their associated scenario parameters.} 
\field{indices}{Scalar operating-point descriptors computed from the power flow, such as load-diversity coefficients, network stress index, voltage margins, loss ratio, aggregate inertia, kinetic energy, total load and generation, and voltage and line-loading statistics. Since the underlying \texttt{powerFlow} and \texttt{mpc} fields are retained, these indices can be re-computed or supplemented with user-defined criteria in post-processing, without re-simulation.} 
\field{graph}{Attributed-graph representation used by the GNN: node and edge feature matrices (with their \texttt{featureNames}, \texttt{featureUnits}, and \texttt{featureGroups}), edge index and adjacency, and node-type information, encoding $\mathcal{G}$.}
\field{validation}{\emph{Steady-state} quality assessment from the screening step: \texttt{status}, 
\texttt{reason}, convergence flag, minimum and maximum load-bus voltages, maximum line loading, reactive-limit-violation indicator, and power-balance error. Characterizes the \emph{pre-fault} point and is \emph{not} the learning target.} \field{transient}{\emph{Dynamic} ground truth from the time-domain simulation (present when the transient layer ran successfully): a \texttt{success} flag; the per-machine trajectories \texttt{time} ($N_t\times1$, s), \texttt{delta} ($N_t\times n_g$, absolute rotor electrical angle, rad) and \texttt{omega} ($N_t\times n_g$, rotor speed, pu), ordered by \texttt{genNames}; the dynamic \texttt{indices}; and the stability \texttt{label}. The \texttt{label} sub-structure holds \texttt{status} (\textsc{stable}/\textsc{unstable}), \texttt{value} ($0$ stable, $1$ unstable), \texttt{maxSeparation} (maximum center-of-inertia rotor-angle separation, deg), \texttt{threshold}, \texttt{margin} $=$ \texttt{threshold} $-$ \texttt{maxSeparation}, and a textual \texttt{criterion}.} 
\field{mpc}{MATPOWER case structure for this scenario (buses, generators, branches, and base MVA), retained so that the exact electrical model underlying the record can be reconstructed or re-solved. Hence, new indicators can be computed by the user for specific analyses.} \noindent\rule{\linewidth}{0.8pt} \smallskip 

\subsection{Reproducibility}
Scenario randomization is fully deterministic: a per-scenario seed is derived from a fixed base seed, 
so the exact dataset can be regenerated or extended without drift. Together with the self-describing, HDF5-based container introduced above, this design makes the dataset compliant with the FAIR principles. The persistent DOI ensures findability and accessibility, while the embedded field names, units, and feature groups make the records interoperable across the MATLAB, Python, and graph-learning tools already noted. Reusability is further supported by the publicly available generation and analysis code and the fixed-seed strategy: future revisions are released as new DOI-referenced versions that preserve earlier releases, and the source code is distributed under the MIT License.

\section{Insights and Notes}
A defining feature of this dataset is that a single record stores the attributed graph, the static machine constants, and the full rotor-angle and speed trajectories together. As a result, all information layers required for topology-aware, physics-based, and hybrid learning approaches are available within the same scenario, a combination that existing open transient-stability resources rarely provide simultaneously. 

Unlike many publicly available transient-stability datasets, each scenario preserves the complete temporal evolution of generator rotor angles and speeds together with the physical quantities required to interpret these dynamics. Consequently, the dataset supports investigations extending beyond binary stability classification, including trajectory prediction, stability-margin estimation, critical-clearing-time analysis, alternative labelling strategies, and dynamic behaviour characterization, without requiring additional simulations.

The dataset can support a broad range of machine-learning and physics-based methodologies, including topology-aware models, time-series prediction approaches, physically constrained learning methods, and hybrid frameworks. Because all information layers are provided within a common data structure and generated under a unified simulation framework, methodological comparisons can be performed without confounding differences in preprocessing, simulation assumptions, or data provenance.

Reproducibility was a primary design objective. All scenarios are generated through a deterministic and resumable pipeline based on fixed random seeds, and the complete generation and analysis code is publicly available. This enables exact regeneration of the released dataset as well as controlled future extensions using compatible procedures.

We reiterate the scope: results characterize operating-point generalization on a single small topology under a specific EMT model, and should not be read as evidence of transfer to larger or structurally different networks.The primary contribution of this dataset is therefore not network scale, but the combination of graph topology, physical machine parameters, complete generator trajectories, deterministic reproducibility, and public accessibility within a single benchmark.  Extending the pipeline to additional network topologies and alternative machine models represents a natural step and is facilitated by the publicly available,  deterministic and resumable generation framework.

\section{Source Code and Scripts}
The full generation and analysis pipeline is implemented in MATLAB and organized into modular stages. Steady-state solutions use MATPOWER~8.1; transient simulations use Simscape Electrical Specialized Power Systems and therefore require a licensed installation to \emph{regenerate} the data. Post-processing and analysis of the saved records, including the composition dashboard and the example-trajectory figures, operate purely on the \texttt{.mat} files and require no Simscape license. Analysis utilities include a label-summary routine, an eight-panel composition dashboard, a representative stable/unstable trajectory plotter, a shared machine-parameter builder, and a stratified nested-split builder. The code is available at~\url{https://github.com/drsup509/ieee9-transient-stability-assessment-graph-dataset}.

\section{Acknowledgments and Interests}

The authors gratefully acknowledge Dr. Nivine Abou Daher and Dr. Ramon Perez Pineda for their insightful review of the project scope.
\\

\noindent H.S., M.D.M., and A.Z. curated and analysed the data, and wrote parts of the manuscript.
\\
\\

\noindent The article authors have declared no conflicts of interest.

\bibliographystyle{IEEEtran}
\bibliography{ref}

\end{document}